\documentclass[aps,prl,showpacs,twocolumn]{revtex4}

\usepackage{graphicx}
\usepackage{longtable}
\usepackage[ansinew]{inputenc}

\begin{document}

\title{Ultraviolet light emission from Si in a scanning tunneling
microscope}

\author{Patrick Schmidt}
\author{Richard Berndt}
\affiliation{Institut für Experimentelle und Angewandte Physik,
Christian-Albrechts-Universität zu Kiel, D-24098 Kiel, Germany}

\author{Mikhail I. Vexler}
\affiliation{A. F. Ioffe Physicotechnical Institute,
Polytechnicheskaya Str.\ 26, 194021 St.-Petersburg, Russia}

\begin{abstract}
Ultraviolet and visible radiation is observed from the contacts of a
scanning tunneling microscope with Si(100) and (111) wafers. This
luminescence relies on the presence of hot electrons in silicon, which
are supplied, at positive bias on $n$- and $p$-type samples, through
the injection from the tip, or, at negative bias on $p$-samples, by
Zener tunneling.  Measured spectra reveal a contribution of direct optical
transitions in Si bulk.  The necessary holes well below the valence
band edge are injected from the tip or generated by Auger processes.
\end{abstract}

\pacs{68.37.Ef, 78.30.Am, 78.60.-b, 73.50.Gr}


\maketitle

The tip of a scanning tunneling microscope (STM) can be used to locally
inject electrons or holes into a sample with atomic-scale precision and to
excite the emission of light \cite{JIM}.  This emission has been observed
and analysed from metals, adsorbed molecules and semiconductors
\cite{KOEN,ALV,CDS,USH,MURA,YOKO,App05}. While on metals the light emission is
excited in the tunneling gap region \cite{RB1}, the fairly intense
luminescence from direct semiconductors has been shown to occur inside
their bulk \cite{ALV,CDS,KOEN}. Recently, tunneling-induced luminescence
from Si(100), an indirect-band material, has been observed and attributed
to inelastic transitions between Si dangling bond states and states
specific to W tips \cite{Aono1,Aono2}.  Very  similar isochromat spectra
were reported from $n$- and $p$-type material independently of the
polarity of the tip-to-sample bias. Besides, the radiation from a STM
contact with Si(111) has been reported and related to a localised plasmon
which was suggested to arise on this Si surface \cite{Downes}. Emission
from the bulk of Si owing to the diffusing carriers was excluded. On the
other hand, luminescence of bulk silicon is a known phenomenon
\cite{HAY56,DUM57}, which is currently being used, e.g., in studies of
metal-oxide-semiconductor (MOS) devices \cite{BUDE2,Vexler}.  A number of
theoretical analyses have addressed the role of direct and phonon-assisted
processes and of impurities \cite{BUDE2,PAV02,VIL95,CAR97}.

Here we undertake an experimental investigation into the STM-induced
luminescence from $n$- and $p$-type Si. For the first time, detailed
luminescence spectra are reported.  We included both (111) and (100)
orientations in our study, in order to vary the Si surface electronic
structure, but the data suggests that this structure does not significantly
affect the spectra. As will be shown further, our results are consistent
with the model of radiative transitions within the bulk Si. In general, the
luminescence of a STM contact has much in common with that of planar MOS
structures. However, while the luminescence studies in MOS devices are
limited to a low bias regime due to the oxide breakdown at elevated
electric fields, this problem is less severe in STM experiments. As a
result, we can observe the emission of ultraviolet photons with energies
exceeding 4 eV\@.

The experiments were carried out in a custom built STM under ultra high
vacuum conditions at ambient temperature \cite{THJ}. Si samples (B doped
\textit{p}-Si(111), 1.5 $\Omega$cm; P doped \textit{n}-Si(111), 56.5
$\Omega$cm; P doped \textit{n}-Si(100), 10 $\Omega$cm) were prepared by
resistive heating. NaOH etched and {\em in vacuo} heated tungsten tips were
used. The photon detection setup was similar to the one reported in Ref.\
\onlinecite{rsi}. A solid angle of 0.6 sr centered at $\angle = 30$$^\circ$
with respect to the surface was detected. Light emission spectra were
recorded on clean, reconstructed (7$\times$7) Si(111) and (2$\times$1)
Si(100) surfaces. In all cases, they were corrected for the wavelength
dependency of the detection sensitivity. The spectra were not noticeably
affected by scanning the tip in a constant current mode. Since tip changes
or surface damage are conveniently diagnosed during scanning we used this
mode for spectroscopy and made sure that tip and surface remained
unaltered.

\begin{figure}[h!]
\includegraphics[width=80mm,clip=]{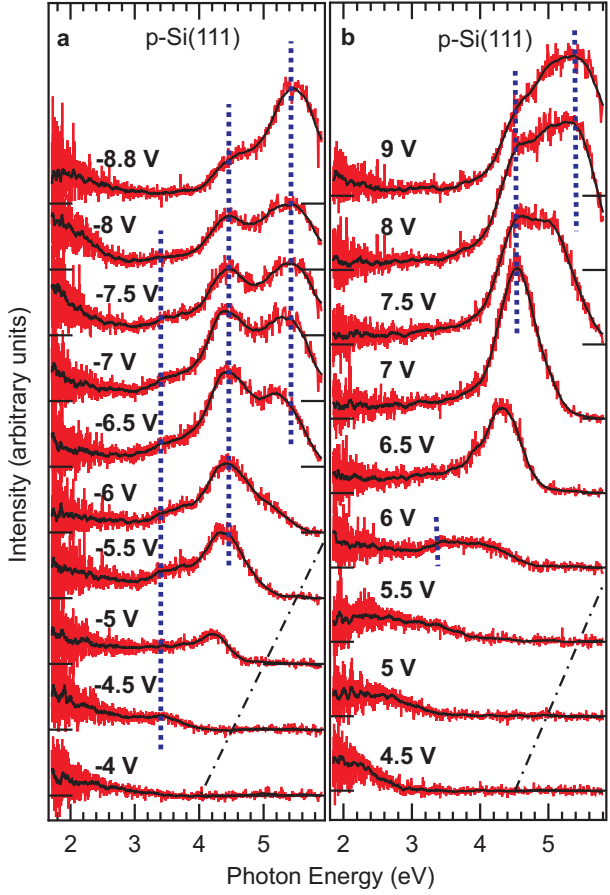}
\caption{Luminescence spectra recorded from \textit{p}-Si(111) at tunneling
current $I = 26.7$ nA for (a) negative and (b) positive sample voltage $V$.
''Noisy'' curves represent raw data acquired with 1340 channels, smoother
curves are obtained by averaging. Photon energies $h\nu=
3.4$, 4.5 and 5.4 eV are marked with vertical dashed lines. Dash-dotted
lines represent the condition $h\nu=eV$.} \label{posi}
\end{figure}

\begin{figure}[t]
\includegraphics[width=80mm,clip=]{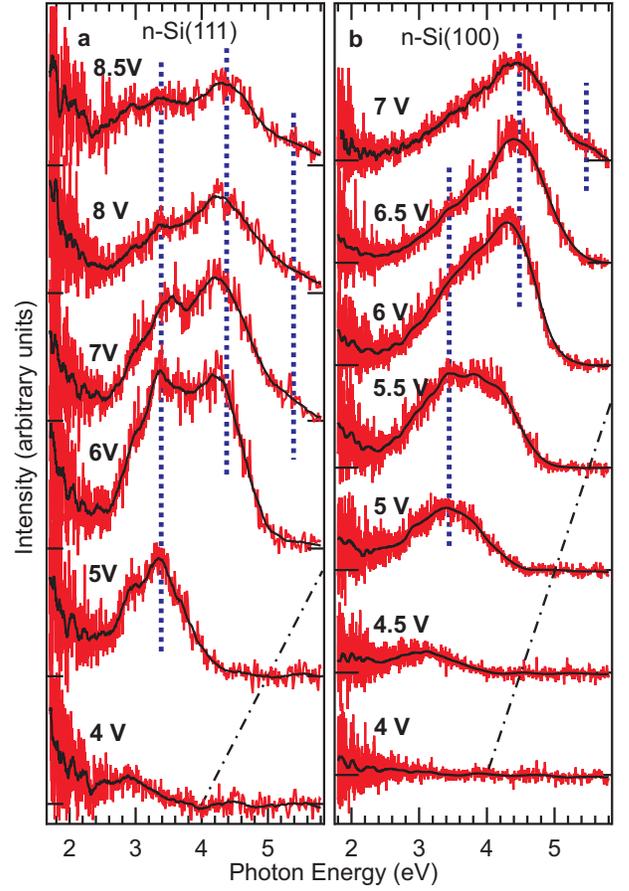}
\caption{Luminescence spectra recorded at positive bias $V$ from (a)
\textit{n}-Si(111) at $I= 9.2$ nA and (b) \textit{n}-Si(100) at $I=10.7$
nA\@.  Dashed and dash-dotted lines defined as in Fig.\ \ref{posi}.}
\label{nega}
\end{figure}

Series of luminescence spectra from \textit{p}-Si(111) recorded at
different sample voltages $V$ are presented in Fig.\ \ref{posi}. For
negative $V$, clear peaks are observed at photon energies $h\nu=4.5$ eV and
5.4 eV\@. An additional structure is discernible at $h\nu=3.4$ eV for $V=
-4.5\ldots -8$ V\@. For positive $V$, the peaks at $h\nu=4.5$ eV and 5.4 eV
were observed too, while at $h\nu=3.4$ eV, enhanced intensity is hardly
discernible except for the $V= 6$ V data. Spectra from \textit{n}-Si (Fig.\
\ref{nega}) exhibit pronounced peaks at $h\nu=3.4$ eV and 4.5 eV, but at
5.4 eV the intensity is only weakly perturbed. All the spectra show an
intensity increase at low photon energies which is consistent with the
luminescence of almost thermalized carriers observed also on Si MOS
capacitors \cite{Yoder}. No radiation was detected from the $n$-Si samples
for $V<0$, unless we previously touched the Si surface with the W tip.

The observed peaks match the energies of direct interband transitions in Si
as determined from reflectance and ellipsometry measurements
\cite{ellipso,reflect,Zucca}.  Three sets of transitions can be
identified \cite{MADEL}:
(a) $\Gamma_{25'} \rightarrow \Gamma_{15}$ and
$L_{3'} \rightarrow L_1$
    at $h\nu \sim 3.4$ eV,
(b) $\Gamma_{25'} \rightarrow \Gamma_{2},
     \Delta_{5'} \rightarrow \Delta_1,
     X_{4'} \rightarrow X_1$,
     $\Sigma_{5'} \rightarrow \Sigma_1$
     at $h\nu \sim 4.5$ eV, and
(c) $L_{3'} \rightarrow L_3$
     at $h\nu \sim 5.4$ eV\@.
Some of these transitions, e.g. $L_{3'} \rightarrow L_3$, involve holes
with energies much lower than that of the valence band edge $E_V$.

In this work we did not intend to perform detailed electrical or optical
simulations considering the exact geometry of a STM contact. Instead, we
restrict ourselves to a qualitative analysis and use models of a
one-di\-men\-sio\-nal (planar) MOS tunnel structure. Such models allow to
estimate the band bending in the semiconductor, the electron and hole
components of the tunneling current through the insulator, the Zener
current within silicon etc.\ and thus capture the essential physics.  Note
that the hole tunneling in a vacuum gap, oppositely to an oxide, occurs
through the upper barrier formed by this gap, as there is no ``insulator
valence band''. Like in a regular MOS structure, the charge states of
depletion, inversion and accumulation can be supported in a STM contact. An
important difference between the planar and real topologies is that the
depletion layer width for the given band bending value will be smaller than
in planar case as can be verified from a calculation for a junction between
a semiconductor and a metal sphere.  When the sphere radius is large, the
flat situation is imitated.

\begin{figure}[h!]
\includegraphics[width=75mm,clip=]{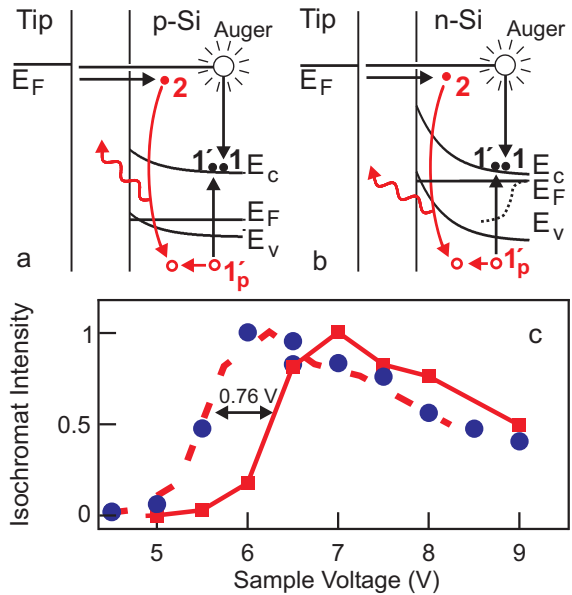}
\caption{
   Band diagrams and relevant processes for STM-induced light emission
   at positive substrate bias $V$.
   (a)
   \textit{p}-Si.  Hot electrons ${\bf 1}$ and ${\bf 2}$ are
   injected from the tip. Electron ${\bf 1}$ undergoes an Auger transition
   creating an electron ${\bf 1'}$ and a hot hole ${\bf 1'_p}$.  Electron
   ${\bf 2}$ radiatively recombines with ${\bf 1'_p}$ which may have drifted
   in the accumulation zone.
   (b)
   \textit{n}-Si. Similar processes occur as in $p$-Si.
   Dotted line indicates the position of quasi Fermi level for holes.
   (c)
   Isochromat spectra for photon energies ($4.28 \pm 0.15$)
   eV from $n$- (dots) and $p$- (squares) Si(111).  Spectra have been
   scaled to identical maximum intensity.  The solid line connects
   \textit{p}-Si(111) data points.  A shift $\Delta =$ 0.76 eV yields
   the dashed line.   } \label{edia}
\end{figure}

\begin{figure}[b]
\includegraphics[width=65mm,clip=]{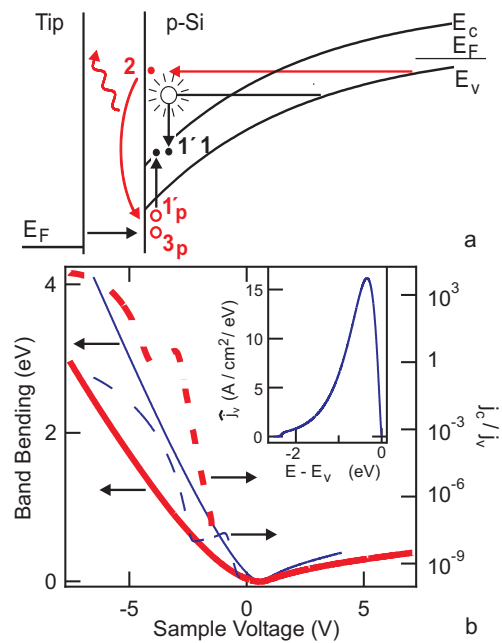}
\caption{
   (a)
   Band diagram and relevant processes for STM-induced light emission from
    \textit{p}-Si at negative bias $V$. Zener tunneling of electrons ${\bf 1}$
and ${\bf 2}$ occurs. ${\bf 1}$ undergoes an Auger transition creating
${\bf 1'}$ and ${\bf 1'_p}$. ${\bf 2}$ radiatively recombines either with
${\bf 1'_p}$ or with a hot hole ${\bf 3_p}$, which tunneled from the tip.
   (b)
   Band bending for a planar W-vacuum-Si junction along with the ratio
$j_c/j_v$ of tunnel components. Calculations were performed for 1 nm (thin
curves) and 2 nm (thick) vacuum gap; dopant concentration is $5 \cdot
10^{18}$ cm$^{-3}$. An energy distribution of the valence band current
(inset, $V = -6.5$ V) shows that hole tunneling some eV below $E_V$ is
significant.} \label{isos}
\end{figure}

Below we use the energy diagrams of the tip-vacuum-semiconductor system, as
if this system were planar, (Figs.\ \ref{edia},\ref{isos}) to illustrate
the processes responsible for light generation.

At positive sample bias, $V>0$ (Figs.\ \ref{edia}a,b), electrons (${\bf
1}$) are injected into Si with energies up to 9 eV above the conduction
band edge of the quasineutral area. At such energies, Auger ionization
processes are the predominant relaxation mechanism with a quantum yield
around one \cite{Brodersen, Vexler2}. The probability for the resulting
hole ${\bf 1'_p}$ to be created significantly below $E_V$ is high
(e.g., for a 4.5 eV electron, the probability of generating a hole with an
energy under $E_V-2\,$eV exceeds $0.2$ \cite{Bude}). A second injected
electron ${\bf 2}$ can then radiatively recombine with ${\bf 1'_p}$. These
optical processes are localized at the depth of about ten nm in Si
\cite{Yoder}, so that a large fraction of emitted photons should reach the
detector without reabsorption.

The above scenario is valid for both \textit{p}- and \textit{n}-type
silicon. However, the energy of the injected electron ${\bf 1}$ with
respect to $E_C$ and, therefore, the threshold for exciting a transition
involving a specific conduction band state for a given bias voltage depends
on doping (Fig.\ \ref{edia}). Experimentally, such a variation is indeed
observed. Fig. \ref{edia}c displays isochromat spectra for the most intense
spectral feature in the photon energy range ($4.28 \pm 0.15$) eV\@.  The
observed shift $\Delta =0.76$ eV between spectra from $n$- and $p$-type Si
agrees with the difference in bulk Fermi energies. This is strong
indication that most of the photons are generated in the bulk material. For
hypothetical near-surface transitions, $\Delta$ should have reflected the
band bending which is rather large and different for $n$- and $p$- Si.

At negative sample voltage, as already mentioned, the light was observed
only for $p$-Si.  The absence of luminescence from $n$-Si at $V<0$ is easy
to understand. The junction operates at forward bias and the band bending
is small. No hot electrons appear in the conduction band of silicon in such
a case.

The origin of luminescence from $p$-Si at $V<0$ can be explained as follows
(Fig.\ \ref{isos}a,b). The junction is reversely biased and strong band
bending occurs. As a consequence, Zener tunneling of electrons from the
valence band is possible; subsequently, these electrons are moving within
the conduction band gaining high energy. Thus hot electrons ${\bf 2}$ and,
via Auger transitions, hot holes ${\bf 1'_p}$ are generated. An alternative
process leading to the appearance of hot holes is tunneling between the tip
and the states in Si underneath the valence band edge (${\bf 3_p}$ in Fig.\
\ref{isos}a, inset to Fig.\ \ref{isos}b). The final step is radiative
recombination of ${\bf 2}$ with ${\bf 1'_p}$ or with ${\bf 3_p}$.

There might be doubts about the validity of a Zener transport mechanism in
our relatively low doped $p$-silicon samples. Indeed, this mechanism is
known from the textbooks to come into the play at fairly high fields for
which high dopant concentrations, say 10$^{18}$-10$^{19}$ cm$^{-3}$, are
required. However, one should not forget that these values refer to the
planar geometry, and the same fields in the depleted region of a
semiconductor may be attained in a STM contact topology even for much more
moderate doping levels.

Fig.\ \ref{isos}b (dashed lines) displays the calculated ratio of
conduction and valence band current densities $j_c/j_v$, where the acceptor
concentration was taken larger than in our samples, to roughly
compensate for the geometry effect. Qualitatively, at high negative bias,
most hot holes are found to be provided by Zener tunneling followed
by Auger processes, while hole injection from the tip becomes more
significant for lower voltages. Note also that for a positive bias $V>0$,
Fig.\ \ref{edia}, the role of valence band tunneling in supplying hot holes
is always minor, since $j_c\gg j_v$.

It is not useless to mention that ordinary MOS structures on highly
doped $p$-Si emit no light for $V<0$, although Zener tunneling certainly
occurs there. This may be due to the intense non-radiative scattering of
hot electrons on defects in the presence of a large impurity concentration.
In a STM contact, due to the difference in a topology, less
highly doped samples can be used, which favors
the observation of luminescence.

In summary, we have reported ultraviolet and visible light from Si(100)
and (111) surfaces in a scanning tunneling microscope.  While electron
injection causes luminescence from $n$- and $p$-type samples, no emission
is detected at negative sample bias from $n$-Si in contrast to previous
reports.  We have presented the first detailed luminescence spectra
revealing a great contribution of direct transitions of hot electrons and
holes. The luminescence from $p$-Si at reverse bias involves Zener
tunneling and injection of holes well below the valence band edge.

Our results, in particular the similarity of the data from (100) and (111)
surfaces, the spectral features observed, and the polarity dependence of
the emission are not consistent with the interpretation of Si light
emission in terms of localised plasmons \cite{Downes}.  They also are at
variance with the interpretation of Refs.\ \onlinecite{Aono1,Aono2} in
terms of an inelastic transition between a specific W tip state and Si
dangling bond states.  It is important to note, however, that the dopant
density of the Si samples used in Refs.\ \onlinecite{Aono1,Aono2} was
substantially higher than in our present work.

We thank P. Johansson, University of Örebro, for many discussions and
for calculations of the electromagnetic response of a tunneling gap between
a W tip and a Si surface. One of the authors (MIV) thanks the A. von Humboldt
foundation for a support of his stay at the TU Braunschweig where part of
this work has been done.

\end{document}